\documentclass[twocolumn,showpacs,amsfonts,aps,prc,nofootinbib,floatfix,%
superscriptaddress]{revtex4}

\usepackage{amsmath}
\usepackage{bm}
\usepackage{graphicx}

\usepackage{epsfig}
\newcommand{\beq}{\begin{equation}}
\newcommand{\eeq}{\end{equation}}
\newcommand{\bea}{\vspace{0.25cm}\begin{eqnarray}}
\newcommand{\eea}{\end{eqnarray}}

\newcommand{\ro}{\mbox{{\boldmath
$\rho$}}}

\newcommand{\db}{{{\bf d}}}
\newcommand{\rb}{\mbox{{\bf
r}}}

\newcommand{\bb}{{{\bf b}}}

\newcommand{\E}{{{\bf E}}}

\newcommand{\Vb}{{{\bf V}}}

\newcommand{\B}{{{\bf B}}}

\def\lsim{\mathrel{\rlap{\lower4pt\hbox{\hskip1pt$\sim$}}
    \raise1pt\hbox{$<$}}}         
\def\gsim{\mathrel{\rlap{\lower4pt\hbox{\hskip1pt$\sim$}}
    \raise1pt\hbox{$>$}}}         

\newcommand{\landau}{L.D.~Landau Institute for Theoretical Physics,
        GSP-1, 117940, Kosygina Str. 2, 117334 Moscow, Russia}

\begin{document}


\title{
Quantum analysis of fluctuations of electromagnetic fields in heavy-ion collisions
}
\date{\today}

\author{B.G.~Zakharov}\affiliation{\landau}

\begin{abstract}
We perform quantum calculations
of fluctuations of the electromagnetic fields in $AA$ collisions
at RHIC and LHC energies. The analysis is based on the 
fluctuation-dissipation theorem.
We find that in the quantum picture the field fluctuations 
are very small. They turn out to be much smaller than the predictions 
of the classical Monte-Carlo simulation with the Woods-Saxon nuclear density.

\end{abstract}
%

\maketitle


{\it Introduction.}
Non-central heavy ion collisions at RHIC and LHC energies
should generate a very strong  magnetic field perpendicular 
to the reaction plane \cite{Kharzeev_B1,Toneev_B1}. 
At the initial moment it can reach the values up to 
$eB\sim 3m_{\pi}^{2}$ for RHIC ($\sqrt{s}=0.2$ TeV)
and a factor of 15 bigger  
for LHC ($\sqrt{s}=2.76$ TeV) 
\cite{Kharzeev_B1,Toneev_B1,Tuchin_B,Z_maxw}.
The presence of the magnetic field may lead to 
charge separation along the magnetic field direction
due to the anomalous current
$\propto \B$ 
(the Chiral Magnetic Effect (CME))
in the quark-gluon plasma (QGP) produced in the initial stage 
of $AA$ collisions
\cite{Kharzeev_B1,Kharzeev_CME_rev}. 
The effect is supported by the experimental data 
on the charged particle correlations
\cite{STAR_CME}.
But the situation remains somewhat unclear due to 
non-CME related background effects
\cite{Kharzeev_report}.

An important issue arising in the context of
the CME and charge separation in
$AA$ collisions concerns fluctuations of the magnetic field. 
They partly destroy the correlation 
between the magnetic field direction and the reaction plane, and can lead 
to reduction of the $\B$-induced observables 
\cite{Liao_MC}.
Fluctuations of the electromagnetic fields 
in $AA$ collisions have been addressed in several studies 
\cite{Skokov_MC,Deng_MC,Liao_MC}
by Monte-Carlo (MC) simulation
with the Woods-Saxon (WS) nuclear distribution
using the classical Lienard-Weichert potentials.
The results of \cite{Skokov_MC,Deng_MC,Liao_MC} show 
that fluctuating proton positions lead to 
considerable event-by-event fluctuations 
of the magnetic field both parallel and perpendicular to the reaction plane.
It would be highly desirable to perform a quantum analysis of the problem.  
Because the classical treatment has no theoretical
justification. Indeed, the dominating contribution
to fluctuations of the electromagnetic fields is connected
with fluctuations of the nuclear dipole moments.
It is well known that the 
nuclear dipole fluctuations
are dominated by the giant dipole resonance (GDR), which is 
a collective excitation closely related to the symmetry energy
of the nuclear matter 
\cite{Greiner,Speth,Trippa}. 
But in the description of nuclei in terms of 
the factorized WS nuclear distribution this collective 
quantum dynamics of the nuclear ground state
is completely ignored. The classical treatment of the electromagnetic field
in the problem of interest may also be inadequate. Because, 
similarly to calculations of the van der Waals forces \cite{Casimir}, 
it becomes invalid at large distances.
\begin{figure} [t]
\vspace{.7cm}
\begin{center}
\epsfig{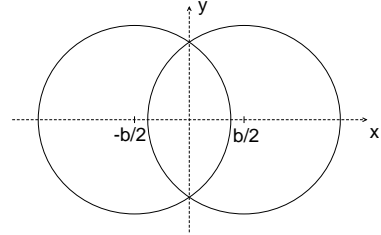}
\end{center}
\vspace{-0.5cm}
\caption[.]
{The transverse plane of a non-central $AA$ collision with the impact
parameter $b$. 
}
\end{figure}

In the present letter we perform a quantum analysis of 
the field fluctuations in $AA$ collisions at RHIC and
LHC energies. Our framework is based on the general formulas of the
fluctuation-dissipation theorem \cite{Callen} 
for the electromagnetic fluctuations in the form given in \cite{LL9}.
This formalism allows to express the field fluctuations in $AA$ 
collisions via the nuclear dipole polarizability.

{\it Theoretical framework.}
We consider the proper time region
$\tau\sim 0.2-1$ fm which is of the most interest from the point
of view of the $\B$-induced effects in the QGP. 
Because in this time region the QGP may be already present, and 
the magnetic field is still significant.
The MC simulation of the electromagnetic fields 
in $AA$ collisions using the retarded Lienard-Weichert potentials
show that almost all contribution to the electromagnetic field
comes from the currents generated by the fast protons in 
the space-time region before intersection of the nuclear disks.
For this reason one can ignore the effect of interaction of the colliding
nuclei on the currents from the participant protons involved in 
the nuclear collision.
In any case, for the RHIC and LHC energies, a possible change of 
the fast quark currents at the time scale $t\lsim 1$ fm
should be very small. 
We make a  reasonable assumption that, similarly to the classical case, 
in the quantum picture
the effect of the electric currents from the fast protons in the stage 
after the nuclear collision is negligible. 
In the present analysis we ignore the electromagnetic fields generated
in the QGP stage by new quarks and antiquarks produced after interaction of the 
Lorentz contracted colliding nuclei \cite{Skokov_B,Tuchin_B,Z_maxw}.
Then the problem is reduced
to evaluating the electromagnetic fields generated by two colliding
nuclei in the ground state.

We consider two colliding nuclei (right moving and left moving) 
with velocities
$\Vb_{R}=(0,0,V)$ and $\Vb_{L}=(0,0,-V)$, and with the impact parameters
$\bb_R=(0,-b/2)$ and $\bb_L=(0,b/2)$ (as shown in Fig.~1).
We take $z_{R,L}=\pm Vt$.
The electromagnetic field from each nucleus
is a sum of the mean field and the fluctuating field
\beq
F^{\mu\nu}=\langle F^{\mu\nu}\rangle+\delta F^{\mu\nu}\,.
\label{eq:10}
\eeq
For each nucleus $\langle \E\rangle$ and $\langle \B\rangle$  are given 
by the Lorentz transformation of its Coulomb field in the nucleus rest frame.
For a nucleus with velocity 
$\Vb=(0,0,V)$ and the impact vector $\bb$  
the mean electric and magnetic fields at $x^{\mu}=(t,\ro,z)$ read
\beq
\langle \E_{T}(t,\ro,z)\rangle =\gamma \frac{
E_{A}(r')(\ro-\bb)}{r'}\,,
\label{eq:20}
\eeq
\beq
\langle E_{z}(t,\ro,z)\rangle=\frac{E_{A}(r')z'}{r'}\,,
\label{eq:30}
\eeq
\beq
\langle \B(t,\ro,z)\rangle=[\Vb\times \langle \E(t,\ro,z)\rangle ]\,.
\label{eq:40}
\eeq
Here $\gamma=1/\sqrt{1-V^{2}}$ is the Lorentz factor, 
$r'^{2}=(\ro-\bb)^{2}+z'^{2}$, $z'=\gamma(z-Vt)$,
and 
\beq
E_{A}(r)=\frac{4\pi}{r^{2}}\int_{0}^{r}d\xi\xi^{2}\rho_{A}(\xi)
\label{eq:50}
\eeq
is the electric field of the nucleus in its rest frame, $\rho_{A}$ is
the nucleus charge density. 
From (\ref{eq:20})--(\ref{eq:50}) one can obtain 
that for two colliding nuclei 
the mean magnetic field at $\rb=0$  has only $y$-component. At 
$t^{2}\gsim (R_{A}^{2}-b^{2}/4)/\gamma^{2}$ (here $R_{A}$ is the 
nucleus radius, and $b$ is assumed to be $<2R_{A}$)
it is approximately
\beq
\langle B_{y}(t,\rb=0)\rangle \approx \frac{\gamma Zeb}{(b^{2}/4+\gamma^{2}V^2t^{2})^{3/2}}\,.
\label{eq:60}
\eeq
At $t\gg R_{A}/\gamma$    
in the region $\rho\ll t\gamma$ $\langle B_{y}(t,\ro,z=0)\rangle$      
takes a simple $\rho$-independent form 
\beq
\langle B_{y}(t,\ro,z=0)\rangle \approx
Zeb/\gamma^{2}t^{3}\,.
\label{eq:70}
\eeq

The contribution of each nucleus to the correlators of the electromagnetic
fields in the lab-frame may be expressed  via 
the correlators in the nucleus rest frame.
For $\gamma\gg 1$ the dominating fluctuations
in the lab-frame are the ones of the transverse  fields.
The transverse components  of the correlators 
of the electric and magnetic fields
can be written as 
\beq
\hspace{-.05cm}\langle \delta E_i \delta E_k \rangle =
\gamma^2\left[\langle \delta E_i \delta E_k \rangle+
V^2e_{3il}e_{3kj}\langle \delta B_l \delta B_j \rangle
\right]_{rf},
\label{eq:80}
\eeq 
\beq
\hspace{-.06cm}
\langle \delta B_i \delta B_k \rangle =
\gamma^2\left[\langle \delta B_i \delta B_k \rangle+
V^2e_{3il}e_{3kj}\langle \delta E_l \delta E_j \rangle
\right]_{rf},
\label{eq:90}
\eeq 
where $i,k$ are the transverse indices and   
the subscript $rf$ on the right-hand side of (\ref{eq:80}), (\ref{eq:90})   
indicates that  the correlators are calculated in the nucleus rest frame.

In calculations of the rest frame correlators
$\langle \delta E_l \delta E_j \rangle$,
$\langle \delta B_i \delta B_k \rangle$
(hereafter we drop the subscript $rf$)
with the help of the FDT
we follow 
the formalism of \cite{LL9} (formulated in the gauge $\delta A^{0}=0$). 
It allows to relate the time Fourier component of the vector potential 
correlator
\bea
\langle \delta A_i(\rb_1)\delta A_k(\rb_2)\rangle_{\omega}=
\frac{1}{2}\int dt e^{i\omega t}
\langle \delta A_i(t,\rb_1)\delta A_k(0,\rb_2)
\nonumber\\
+\delta A_k(0,\rb_2)\delta A_i(t,\rb_1)\rangle\hspace{2cm}
\label{eq:100}
\eea
and that of the retarded Green's function
\bea
D_{ik}(\omega,\rb_1,\rb_2)=-i\int dt e^{i\omega t}
\theta(t)
\langle \delta A_i(t,\rb_1)\delta A_k(0,\rb_2)
\nonumber\\
-\delta A_k(0,\rb_2)A_i(t,\rb_1)
\rangle\,.\hspace{2cm}
\label{eq:110}
\eea
In the zero temperature limit the FDT relation between (\ref{eq:100}) 
and (\ref{eq:110}) reads \cite{LL9} 
\bea
\!\!\langle \delta A_i(\rb_1)\delta A_k(\rb_2)\rangle_{\omega}\!=\!
-\mbox{sign}(\omega)
\mbox{Im} D_{ik}(\omega,\rb_1,\rb_2).
\label{eq:120}
\eea
The time Fourier components of 
the electromagnetic field correlators in terms of that for the 
the vector potential correlator (\ref{eq:100}) are given by
\beq
\langle \delta E_i(\rb_1)\delta E_k(\rb_2)\rangle_{\omega}=
\omega^{2}\langle \delta A_i(\rb_1)\delta A_k(\rb_2)\rangle_{\omega}\,,
\label{eq:130}
\eeq
\beq
\hspace{-.18cm}
\langle \delta B_i(\rb_1)\delta B_k(\rb_2)\rangle_{\omega}=
\mbox{rot}^{(1)}_{il}
\mbox{rot}^{(2)}_{kj}
\langle \delta A_l(\rb_1)\delta A_j(\rb_2)\rangle_{\omega}.
\label{eq:140}
\eeq
The same point field correlators that we need read 
\beq
\hspace{-.07cm}\langle \delta E_i(t,\rb)\delta E_k(t,\rb)\rangle
=\frac{1}{2\pi}\int_{-\infty}^{\infty}d\omega
\langle \delta E_i(\rb)\delta E_k(\rb)\rangle_{\omega}\,,
\label{eq:150}
\eeq
\beq
\hspace{-.08cm}\langle \delta B_i(t,\rb)\delta B_k(t,\rb)\rangle
=\frac{1}{2\pi}\int_{-\infty}^{\infty}d\omega
\langle \delta B_i(\rb)\delta B_k(\rb)\rangle_{\omega}\,.
\label{eq:160}
\eeq

In the time region of interest $t \gsim 0.2$ fm 
(in the lab-frame) 
in (\ref{eq:150}), (\ref{eq:160}) for each nucleus
the distance between the observation
point $\rb$ and  the center of the nucleus (in its rest frame)
is much bigger than $R_A$.
In this regime one may treat each nucleus
as a point like dipole described by the dipole polarizability 
$\alpha_{ik}(\omega)$ (in the sense of the fluctuating field components).
In the formalism of \cite{LL9} the field fluctuations are described 
by correction to the retarded Green's function
proportional to the dipole polarizability.
The retarded Green's function coincides with the Green's function of 
Maxwell's equation \cite{LL9}.
In the presence of the point like dipole at $\rb=\rb_A$
the equation determining the retarded Green's function reads
\bea
\left[\frac{\partial^2}{\partial x_i\partial_l}-\delta_{il}\triangle
-\delta_{il}\omega^2
-4\pi\omega^2\alpha_{il}(\omega)\delta(\rb-\rb_A)\right]
\nonumber\\
\times D_{lk}(\omega,\rb,\rb')
=-4\pi\delta_{ik}\delta(\rb-\rb')\,.\hspace{1cm}
\label{eq:170}
\eea
The correction to $D_{ik}$ due to $\alpha_{ik}$
reads \cite{LL9}
\bea
\Delta  D_{ik}(\omega,\rb_1,\rb_2)=-
\omega^2 D_{il}^{v}(\omega,\rb_1,\rb_A)\alpha_{lm}(\omega)\nonumber\\
\times D_{mk}^{v}(\omega,\rb_A,\rb_2)\,.
\label{eq:180}
\eea
Here $D_{ik}^{v}$ is the vacuum Green's function that is given by
\beq
D_{ik}^{v}(\omega,\rb_1,\rb_2)=\delta_{ik}D_1(\omega,r)+
\frac{x_ix_k}{r^2}D_2(\omega,r)\,,
\label{eq:190}
\eeq
where $\rb=\rb_1-\rb_2$, and
\beq
D_1(\omega,r)=-\frac{e^{i\omega r}}{r}\left(1+\frac{i}{\omega r}
-\frac{1}{\omega^2 r^2}\right)\,,
\label{eq:200}
\eeq
\beq
D_2(\omega,r)=\frac{e^{i\omega r}}{r}\left(1+\frac{3i}{\omega r}
-\frac{3}{\omega^2 r^2}\right)\,.
\label{eq:210}
\eeq

For spherical nuclei the polarizability tensor can be written
as $\alpha_{ik}(\omega)=\delta_{ik}\alpha(\omega)$. 
$\alpha(\omega)$ is an analytical function 
of $\omega$ in the upper half-plane \cite{LL4}. It satisfies the 
relation $\alpha^{*}(-\omega^{*})=\alpha(\omega)$ \cite{LL4} It means that
on the upper imaginary axis $\alpha(\omega)$ is real.
Using this fact, from Eqs. (\ref{eq:180})--(\ref{eq:210}) one can obtain 
for the rest frame
field correlators (we take $\rb_A=0$)
\beq
\langle \delta E_i(t,\rb)\delta E_k(t,\rb)\rangle=
\delta_{ik}J_1(r)+\frac{x_i x_k}{r^2}J_2(r)\,,
\label{eq:220}
\eeq
\beq
\langle \delta B_i(t,\rb)\delta B_k(t,\rb)\rangle=
\left(\delta_{ik}-\frac{x_ix_k}{r^2}\right)J_3(r)\,,
\label{eq:230}
\eeq
where
\beq
J_1=\frac{1}{2\pi r^7}\left[ I_0+I_1+\frac{3}{4}I_2+\frac{1}{4}I_3
+\frac{1}{16}I_4\right]\,,
\label{eq:240}
\eeq
\beq
J_2=\frac{1}{2\pi r^7}\left[ 3I_0+3I_1+\frac{1}{4}I_2-\frac{1}{4}I_3
-\frac{1}{16}I_4\right]\,,
\label{eq:250}
\eeq
\beq
J_3=-\frac{1}{8\pi r^7}\left[ I_2+I_3+\frac{1}{4}I_4\right]\,,
\label{eq:260}
\eeq
\beq
I_n=\int_{0}^{\infty} d\xi \xi^n e^{-\xi}\alpha\left(\frac{i\xi}{2
  r}\right)\,.
\label{eq:270}
\eeq
These formulas allow to express the fluctuations of the electromagnetic
fields of each nuclei via the dipole polarizability $\alpha(\omega)$. 

{\it Parametrization of the dipole polarizability.}
The function $\alpha(\omega)$ reads \cite{LL4,Migdal_GDR}
\beq
\alpha(\omega)=\frac{1}{3}\sum_{s}\left[
\frac{|\langle 0|\db|s\rangle|^2 }
{\omega_{s0}-\omega -i\delta}+
\frac{|\langle 0|\db|s\rangle|^2 }
{\omega_{s0}+\omega +i\delta}\right]\,,
\label{eq:280}
\eeq
where $\db$ is the dipole operator
\beq
\db=e\frac{N}{A}\sum_p \rb_p-e\frac{Z}{A}\sum_n \rb_n\,. 
\label{eq:290}
\eeq
At $\omega>0$ the dipole polarizability tensor coincides with
the photon scattering tensor \cite{LL4}. This allows to express 
the imaginary part of $\alpha(\omega)$ 
in terms of the dipole photoabsorption cross section as
\beq
\sigma_{abs}(\omega)=4\pi\omega\mbox{Im}\alpha(\omega)\,.
\label{eq:300}
\eeq
For heavy nuclei the dipole strength 
is dominated by the GDR
\cite{Greiner,GDR_RMP75,Speth}. 
It appears as a broad peak in the $\sigma_{abs}$
with a mean energy $\sim 14$ MeV \cite{GDR_RMP75}. 
We parametrize the dipole polarizability
for $^{197}$Au and $^{208}$Pb nuclei
by a single GDR state   
\beq
\alpha(\omega)=
c 
\left[
\frac{1}{\omega_{10}-\omega -i\Gamma/2}+
\frac{1}{\omega_{10}+\omega +i\Gamma/2}
\right]\,.
\label{eq:310}
\eeq
By fitting the data on the photoabsorption
cross section from \cite{GDR_Au} for $^{197}$Au and from \cite{GDR_Pb} for 
$^{208}$Pb 
we obtained the following values of the
parameters: $\omega_{10}\approx 13.6$ MeV, $\Gamma\approx 4.38$ MeV, 
$c\approx 18.2$ GeV$^{-2}$ for $^{197}$Au, and
$\omega_{10}\approx 13.3$ MeV, $\Gamma\approx 3.72$ MeV, $c\approx 18.93$
Gev$^{-2}$ for $^{208}$Pb. 
\begin{figure} [t]
\vspace{.7cm}
\begin{center}
\epsfig{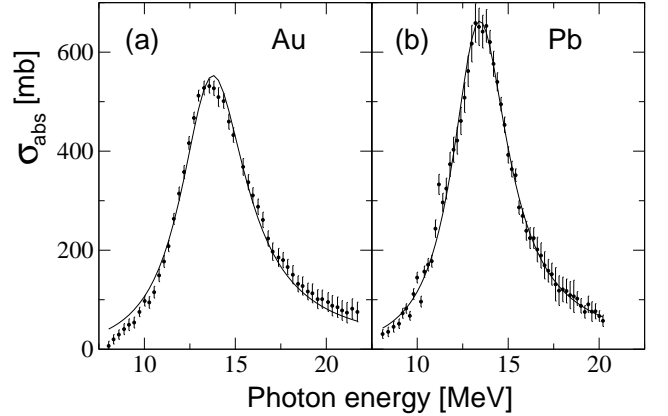}
\end{center}
\vspace{-0.5cm}
\caption[.]
{Fit of the photoabsorption cross section in the GDR region 
to the experimental data for  $^{197}$Au and $^{208}$Pb targets.
The data are from Refs. \cite{GDR_Au} and \cite{GDR_Pb}, respectively. } 
\end{figure}
Fig.~2 illustrates the quality of our fit.

{\it Results and Discussion.} 
The fluctuations of the electromagnetic fields occur due to the fluctuations
of the nuclear dipole moment. For this reason, 
it is interesting to begin with comparison of the 
dipole moment fluctuations in the quantum and the classical models.
In the quantum model from (\ref{eq:280}), (\ref{eq:310}) 
one can obtain 
\beq
\langle 0 |\db^2|0\rangle=\frac{3}{\pi}
\int_{0}^{\infty}d\omega \mbox{Im}\alpha(\omega)
=\frac{6c}{\pi}
\mbox{arctg}\left(2\omega_{10}/\Gamma\right)\,.
\label{eq:320}
\eeq
This formula with parameters fitted to the data on
$\sigma_{abs}$ gives $\langle 0 |\db^2|0\rangle\approx 1.91$ fm$^2$
and $\langle 0 |\db^2|0\rangle\approx 2.02$ fm$^2$  for $^{197}$Au 
and $^{208}$Pb nuclei, respectively.
The classical MC calculation with the WS
nuclear density gives for these nuclei the values 
$\langle \db^2 \rangle\approx 9.89$ fm$^2$
and $\langle \db^2 \rangle\approx10.39$ fm$^2$.
One sees that 
the classical
treatment overestimates the dipole moment squared by a factor of $\sim 5$.
\begin{figure} [t]
\vspace{.7cm}
\begin{center}
\epsfig{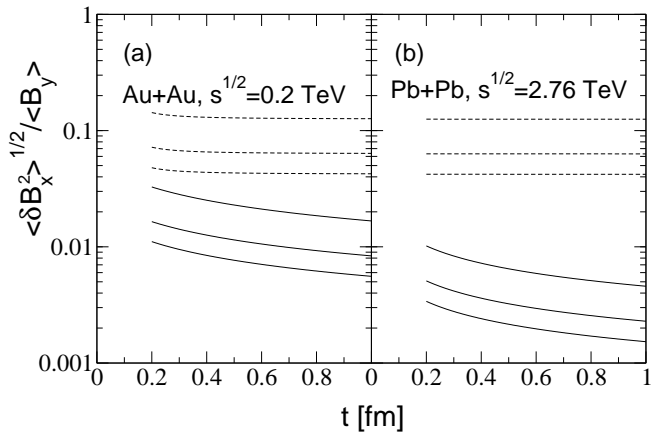}
\end{center}
\vspace{-0.5cm}
\caption[.]
{The $t$-dependence of the ratio 
$\langle \delta B_x^2\rangle^{1/2}/\langle B_y\rangle$ 
at $\rb=0$ for Au+Au collisions at $\sqrt{s}=0.2$ TeV (left)
and for Pb+Pb collisions at $\sqrt{s}=2.76$ TeV
for different impact parameters $b=3$, $6$ and $9$ fm
(from top to bottom).
Solid lines are for quantum calculations, dashed
lines for classical MC calculations with
the WS nuclear density.}
\end{figure}

The fluctuations of the direction of the magnetic field at the 
center of the plasma fireball are dominated
by the fluctuations of the component $B_x$ that vanishes without fluctuations.
In Fig.~3 we confront our quantum and classical results for $t$-dependence 
of the ratio $\langle \delta B_x^2\rangle^{1/2}/\langle B_y\rangle$ 
at $x=y=0$. This ratio gives the typical
angle between the magnetic field and the perpendicular to the reaction plane.
The results are shown for the impact parameters $b=3,$ $6$ and $9$ fm
for Au+Au collisions at $\sqrt{s}=0.2$ TeV and Pb+Pb collisions at
$\sqrt{s}=2.76$ TeV.
One can see that the quantum calculations give 
$\langle \delta B_x^2\rangle^{1/2}/\langle B_y\rangle$
smaller than
the classical ones by a factor of $\sim 5-8$ for RHIC and 
by a factor of $\sim 13-27$ for
LHC. This difference results from both the reduction of the nuclear 
dipole moment fluctuations in the quantum picture and from the quantum effects
for the electromagnetic field (that are especially important for the LHC
energy). We have presented the results for the magnetic field. The results
for fluctuations of the transverse electric field at $\rb=0$ 
are close to that for the magnetic field (recall that $\langle E_{x,y}\rangle=0$
at $\rb=0$).

Thus we see that in the quantum picture 
both for RHIC and LHC fluctuations of 
the direction of the magnetic field relative to the reaction plane are
very small.
Of course, 
in event-by-event measurements the reaction plane itself cannot be
determined exactly. Experimentally the orientation of the reaction 
plane is extracted from the elliptic flow in the particle
distribution
\cite{Ollitrault_v2,Voloshin_v2}, and it fluctuates around the real 
reaction plane. This plane extracted from the data is often called 
the participant plane.
In the hydrodynamical
picture of the QGP evolution calculations of the fluctuations
of the direction of the magnetic field to the 
participant
plane require a joint analysis of the field fluctuations
and of the fluctuations of the initial entropy deposition.
The latter control the fluctuations of the orientation of 
the participant plane. 
The initial entropy deposition is sensitive to the long range 
fluctuations in the nuclear density. 
One of the types
of the collective nuclear modes that can be important is the 
fluctuations related to the GDR. But another collective 
modes such as the giant monopole resonance (corresponding to spherically 
symmetric nuclear oscillations) and the giant quadruple resonance 
\cite{Greiner} 
may also be important for  the participant plane fluctuations.
Our analysis shows that for the dipole mode the classical treatment
based on the MC simulation 
with the WS
nuclear density overestimates the fluctuations.
It would be of great interest to clarify the situation 
for other collective modes.
In particular, this is of great interest for the event-by-event hydrodynamic
simulations of $AA$ collision. 
All this, however, is far beyond of the scope of the present work.

{\it Conclusion}. In this work within the FDT formalism of \cite{LL9} 
we have performed a quantum
analysis of fluctuations of the electromagnetic field in $AA$ collisions
at RHIC and LHC energies. 
Our quantum calculations show that the field fluctuations are very small,
and  they practically do not affect the direction of 
the magnetic field as compared to the mean field classical predictions.
By confronting our quantum results with that from 
the classical MC simulation with the WS nuclear 
distribution,
we have demonstrated that the classical picture overestimates strongly 
the field fluctuations.
Our results are in contradiction with the conclusion of the recent analysis
\cite{Tuchin_quantum}, where it was argued that the quantum diffusion
of the protons may be very important. However, the analysis 
\cite{Tuchin_quantum} is performed for free particles, and the results
are inapplicable directly to nuclei in the ground state.
  
\begin{acknowledgments}
This work is supported by the RScF grant 16-12-10151.
\end{acknowledgments}

\end{document}